\documentclass[5p]{elsarticle}
\usepackage{amssymb}
\usepackage{amsmath}
\usepackage[dvips]{color}

\newcommand{\btab}{\begin{tabbing}}
\newcommand{\etab}{\end{tabbing}}

\newcommand{\beqn}{\begin{equation}}
\newcommand{\eeqn}{\end{equation}}
\newcommand{\barr}[1]{\begin{array}{#1}}
\newcommand{\earr}{\end{array}}
\newcommand{\beqna}{\begin{eqnarray}}
\newcommand{\eeqna}{\end{eqnarray}}
\newcommand{\btablec}{\begin{table} \begin{center}}
\newcommand{\etablec}{\end{center} \end{table}}

\newcommand{\gapproxeq}{\lower.7ex\hbox{$\;\stackrel{\textstyle>}
{\sim}\;$}}

\begin{document}
\begin{frontmatter}

\title{The strangeness form factors of the proton \\within nonrelativistic constituent quark model revisited}

\author{Alvin Kiswandhi}
\author{Hao-Chun Lee}
\author{Shin Nan Yang}

\address{Department of Physics and Center for Theoretical Sciences, \\National Taiwan University,
Taipei 10617, Taiwan}

\begin{abstract}

We reexamine, within the nonrelativistic constituent quark model
(NRCQM), a recent claim that  the current data on the strangeness
form factors indicates that the $uuds\bar s$ component in the proton
is such that the $uuds$ subsystem has the mixed spatial symmetry
$[31]_X$ and flavor spin symmetry $[4]_{FS}[22]_F[22]_S$, with $\bar
s$ in $S$ state (configuration I). We find this claim to be invalid
if corrected expressions for the contributions of the transition
current to $G^s_A$ and $G^s_E$ are used. We show that, instead, it
is the lowest-lying $uuds\bar s$ configuration with $uuds$ subsystem
of completely symmetric spatial symmetry $[4]_X$ and flavor spin
symmetry $[4]_{FS}[22]_F[22]_S$, with $\bar s$ in $P$ state
(configuration II), which could account for the empirical signs of
all form factors $G^s_E, G^s_M,$ and $G^s_A$. Further, we find that
removing the center-of-mass motion of the clusters will considerably
enhance the contributions of the transition current. We also demonstrate
that it is possible to give a reasonable description of the existing
form factors data with a tiny probability $P_{s\bar s}=0.025\%$ for
the $uuds\bar s$ component. We further see that with a small admixture
of configuration I , the agreement of our prediction with the data for $G_A^s$ at low-$q^2$
region can be markedly improved. We find that without removing CM
motion, $P_{s\bar s}$ would be overestimated by about a factor of
four in the case when transition current dominates. We also explore
the consequence of a recent estimate reached from analyzing the existing
data on $\bar d -\bar u$ , $s +\bar s$ , and $\bar{u} + \bar{d} - s -\bar{s}$, that $P_{s\bar s}$ lies between $2.4-2.9\%$.
It would lead to a large size for the five-quark system and a small
bump in both $G^s_E+\eta G^s_M$ and $G^s_E$ in the region of $q^2\le
0.1 \,\textrm{GeV}^2$ within the considered model.

\end{abstract}


\end{frontmatter}

The first indications of possible existence of strangeness content
in the proton  came from deep-inelastic muon scattering, elastic
neutrino-proton scattering, and analyses of $\pi N$ $\sigma$-term
\cite{Beck01}. Many other observables were later suggested,
including excess $\phi$ production in $p\bar p$ annihilation
\cite{Amsler98}, double polarizations in photo- and
electroproduction of $\phi$ meson \cite{TOY}, and asymmetry in
scattering of longitudinally polarized  electrons from polarized
targets. While the measurement of double polarization in $\phi$
photoproduction is being  pursued with the development of polarized
HD target by LEPS at SPring-8 \cite{Ohta11},  four vigorous
experimental programs SAMPLE \cite{SAMPLE}, HAPPEx \cite{HAPPEX}, A4
\cite{A4}, and G0 \cite{G0} have already been undertaken  to measure
parity-violating asymmetry of polarized electron-proton scattering
in order to extract proton strangeness electromagnetic form factors.

On the theoretical side, lattice QCD remains the only theoretical
method which could provide a reliable determination of the strangeness
form factors from the first principle. For example, a recent low-mass
quenched lattice QCD simulation, with the use of chiral
extrapolation technique and the assumption of charge symmetry, gives
$\mu_S=(-0.046\pm 0.019)\mu_N$ \cite{Leinweber05} and $G_E^s(Q^2=0.1~
\textrm{GeV}^2)=-0.009\pm 0.005\pm 0.003\pm 0.027$ \cite{Leinweber06}. More
recent LQCD efforts can be found in Ref. \cite{LQCD10}.
 Nevertheless, a study of this intriguing question within hadron models
could still provide invaluable insight concerning the underlying
quark structure, e.g., how the strange quarks are arranged inside
proton.

Recently, Zou and Riska  \cite{ZR05} considered  the possible
low-lying configurations of the $uuds\bar s$ component of the proton
within a constituent quark model. It was concluded that the
empirical indications of a positive strangeness magnetic moment of
the proton \cite{SAMPLE,HAPPEX,A4} suggest that the dominant $s\bar
s$ configuration in the proton would have the $\bar s$ in the ground
state and   $uuds$ system  in the $P$ state. It would lead to an
interesting implication that the $qqqq\bar q$ components in the
proton would mainly be in colored quark cluster configurations
rather than in "meson cloud" configurations as commonly
perceived. The calculation of Ref. \cite{ZR05} was later extended to
the evaluation of proton strangeness form factors \cite{RZ06,An06}.

The calculation of Refs. \cite{ZR05,RZ06,An06} did not remove the center-of-mass (CM)
motion of the quark clusters. That could conceivably affect the
estimate of the probability of the $s\bar s$ configuration from the
measured strangeness magnetic moment considerably, as proposed in
Ref. \cite{ZR05}. Accordingly, we set forth to reexamine the
problem with the removal of the CM motion of the clusters. In the
process, we obtained results which differ substantially from those
presented in Refs. \cite{ZR05,RZ06,An06} where the CM motion was not
removed.

The configurations of the $uuds\bar s$ component in the proton
considered in Refs. \cite{ZR05,RZ06,An06} are all of (4,1) clustering
type in that either four quarks $uuds$ would be in $P$ state with $\bar
s$ in $S$ state (configuration I) or $uuds$ in $S$ state while $\bar s$
in $P$ state (configuration II), respectively, in order to ensure
$uuds\bar s$ as a whole has  positive parity as proton. The symmetry
of the spatial state of the four-quark system in configurations I
and II would then be of $[31]_X$ and completely symmetric $[4]_X$,
respectively. Within the harmonic oscillator constituent quark
model, these two configurations are degenerate and of the lowest
energy. However, the degeneracy is lifted by the color hyperfine
quark-quark interaction as shown in Ref. \cite{ZR05}. After the
splitting, the states of the lowest energy in configurations I and
II for $uuds$ cluster would have the flavor and spin state symmetry
of $[4]_{FS}[22]_F[22]_S$ and $[31]_{FS}[211]_F[22]_S$, respectively
\cite{ZR05}. We will focus only on these two states of the lowest
energy in this study.

The calculation of the strangeness form factors $G^s_E, G^s_M,$ and
$G^s_A$ within a harmonic oscillator constituent quark model is
straightforward. However, the evaluation of the contribution of the transition
current is involved if the CM motion is to be removed.

We first compare our results for the form factors with those
obtained in Ref. \cite{RZ06} where the CM motion is not removed. In
order to facilitate the comparison, we will follow the notations and
conventions of Refs. \cite{RZ06,An06} as much as possible, unless
otherwise specified.

There are already some differences between our results and those
given in Ref. \cite{RZ06} for the simple case of diagonal matrix
elements of the vector and axial vector current operators. Besides
the factor of $\sqrt{1+q^2/4m_s^2}$ introduced in the denominators
for $G_E^s$ and $G^s_M$ to account for relativistic effects, we have
not been able to reproduce the factor of $(1-q^2/18\omega_5^2)$ (we
use $\omega_5$ to denote  the harmonic oscillator parameter
associated with the $uuds\bar s$ component so our $\omega_5$ is
$\omega$ used in \cite{RZ06}) which appears in the expression for
$G^s_M$. However, those differences are numerically insignificant in
the low-$q^2$ region where the nonrelativistic model calculations
are, at best, expected to be valid.

For the contribution of transition current, i.e., the non-diagonal
matrix elements of the currents between $3q$ and $5q$ states, there
are three discrepancies between our results and those of Ref.
\cite{RZ06}. First, there is a sign difference for $G_E^s$, as well
as that our result for $G_M^s$ is $\sqrt{3}$ times larger than that
presented in Ref. \cite{RZ06}, which arises from a simple evaluation
of the color matrix element \cite{An11}. The last and the most
serious difference lies within $G_A^s$ in which we obtain the
following expression for the contribution of transition current to
$G_A^s$ \beqna G_A^{s,\textit{ND}}(q^2) = \delta
\frac{g_s}{g_p}C_{35}\frac{2\alpha_s}{m_s} e^{-q^2/4\omega_5^2}
\sqrt{P_{uud}P_{s\bar{s}}}, \label{IGAwCM} \eeqna where the
superscript $\it{ND}$ specifies that it is a non-diagonal matrix
element. The parameters $\omega_3, P_{uud}$ and $\omega_5,
P_{s\bar{s}}$ denote the usual oscillator parameters and
probabilities, respectively, of the $uud$ and $uuds\bar s$
configurations in the proton, and
$C_{35}\equiv(2\omega_3\omega_5/(\omega_3^2+\omega_5^2))^{9/2}$,
while $\delta$ denotes the relative phase between the $uud$ and
$uuds\bar s$ components of the wave functions in the proton.   Eq.
(\ref{IGAwCM}) approaches constant at $q^2=0$, while the
corresponding result of Ref. \cite{RZ06} contains a factor of $q^2$
and hence vanishes at $q^2=0$. Since, as we see later, the
contributions of non-diagonal matrix elements would dominate over
the diagonal ones for all reasonable choices of $\omega_3$ and
$\omega_5$, our result of Eq. (\ref{IGAwCM}) would lead to a
consequence that $G^s_M$ and $G^s_A$ would be of the same sign,
irrespective of the choice of phase $\delta$, at low-$q^2$ region
which is in contradiction with the existing experimental data.  It
would then exclude the possibility that configuration I with four
quarks $uuds$ in $P$ state and $\bar s$ in $S$ state could be the
dominant configuration for the $uuds\bar s$, as concluded in Ref.
\cite{RZ06}. We have carried out a direct numerical six-dimensional
integration to verify that results  agree with the analytical
expression of  Eq. (\ref{IGAwCM}) numerically.

When the center-of-mass motion of the five-quark $(5q)$ cluster is
removed, we obtain the following results, in the case of
configuration I, for the contributions of the diagonal matrix
elements of the current to the proton strangeness form factors.
\beqna G_E^{s,D}(q^2) &=& - \frac{q^2}{24\omega_5^2}
e^{-q^2/5\omega_5^2} P_{s\bar{s}}, \label{IGEwoCM} \\
G_M^{s,D}(q^2) &=&  \frac{m_p}{2m_s} e^{-q^2/5\omega_5^2} P_{s\bar{s}}, \label{IGMwoCM}  \\
G_A^{s,D}(q^2) &=& -\ \frac{1}{3} e^{-q^2/5\omega_5^2} P_{s\bar{s}},
\label{IGAwoCM} \eeqna where the superscript $D$ indicates that they
are the  diagonal matrix elements.  Without the removal of the CM
motion of the $5q$ cluster, the Gaussian factor
$e^{-q^2/5\omega_5^2}$ in Eqs. (\ref{IGEwoCM}-\ref{IGAwoCM}) would
become $e^{-q^2/4\omega_5^2}$. This means that the removal of CM
motion causes the form factors to decrease more slowly. Our results for the
transition $(3q-5q)$ and $(5q-3q)$ contributions to the strangeness form
factors, when the CM of $3q$ and $5q$ clusters are removed read as

\beqna
G_E^{s,\textit{ND}}(q^2) &=& \delta C_{35}^{2/3}\frac{2\cdot15^{3/4}}{9\sqrt{3}}\frac{q^2}{m_s\omega_5} e^{-4q^2/15\omega_5^2} \nonumber\\
&\times&\sqrt{P_{uud}P_{s\bar s}}\,, \label{IGENDwoCM}\\
G_M^{s,\textit{ND}}(q^2) &=& \delta C_{35}^{2/3}\frac{2\cdot15^{3/4}}{9\sqrt{3}}\frac{4m_p}{\omega_5} e^{-4q^2/15\omega_5^2} \nonumber\\
&\times&\sqrt{P_{uud}  P_{s\bar s}}\,, \label{IGMNDwoCM} \\
G_A^{s,\textit{ND}}(q^2) &=& \delta C_{35}^{2/3}\frac{2\cdot15^{3/4}}{9\sqrt{3}}\frac{3\omega_5}{m_s}e^{-4q^2/15\omega_5^2} \nonumber\\
&\times&\sqrt{ P_{uud}  P_{s\bar s}}\,. \label{IGANDwoCM}\eeqna One
observes that the exponents of the Gaussians in Eqs.
(\ref{IGEwoCM}-\ref{IGAwoCM}) and Eqs.
(\ref{IGENDwoCM}-\ref{IGANDwoCM}) are different. This is expected
since the center of masses of the $3q$ and $5q$ clusters are also different. It is
interesting to see that the transition current contributions drop
faster than the diagonal ones. Furthermore, it is seen that
$G_M^{s,ND}$ and $G_A^{s,ND}$ are of the same sign as is the case
before the removal of the CM motion, independent of the relative
phase between the wavefunctions of $3q$ and $5q$ components.
Consequently, as long as the transition current contributions
dominate over the direct terms, then the configuration with
$\bar s$ in $S$ state cannot be the dominant configuration for
$uuds\bar s$ component.

We next study configuration II which is degenerate with the lowest
energy configuration I before being lifted by the color hyperfine
quark-quark interaction. In configuration II, $uuds$ cluster is in
$S$ state while $\bar s$ is in $P$ state. Only the results with the
removal of CM motion will be presented. The direct current gives
rise to the following contributions,

\beqna
G_E^{s,D}(q^2) &=& \frac{q^2}{8\omega_5^2} e^{-q^2/5\omega_5^2} P_{s\bar{s}} \label{IIGEDwoCM}\\
G_M^{s,D}(q^2) &=& \frac{m_p}{m_s} \left(\frac{-1}{6} - \frac{2q^2}{15\omega_5^2}\right) e^{-q^2/5\omega_5^2} P_{s\bar{s}} \label{IIGMDwoCM} \\
G_A^{s,D}(q^2) &=& \left(\frac{-1}{3} +
\frac{2q^2}{15\omega_5^2}\right) e^{-q^2/5\omega_5^2} P_{s\bar{s}}.
\label{IIGADwoCM} \eeqna The results for the transition current
matrix elements between ($3q-5q$ and $5q-3q$) read as follows,
\beqna
G_E^{s,\textit{ND}}(q^2) &=& \delta C_{35}^{2/3}\left(\frac{2}{5}\right)^{1/2}\left(\frac{5}{3}\right)^{3/4}\frac{q^2}{m_s\omega_5} \nonumber\\
&\times&e^{-4q^2/15\omega_5^2}\sqrt{P_{uud} P_{s\bar{s}}}, \label{IIGENDwoCM}\\
G_M^{s,\textit{ND}}(q^2) &=& \delta C_{35}^{2/3}\left(\frac{2}{5}\right)^{1/2}\left(\frac{5}{3}\right)^{3/4}\frac{4m_p}{\omega_5}\nonumber\\
&\times&e^{-4q^2/15\omega_5^2}\sqrt{P_{uud} P_{s\bar{s}}},  \label{IIGMNDwoCM} \\
G_A^{s,\textit{ND}}(q^2) &=& -\delta C_{35}^{2/3}\left(\frac{2}{5}\right)^{1/2}\left(\frac{5}{3}\right)^{3/4}\frac{5\omega_5}{m_s}\nonumber \\
&\times&e^{-4q^2/15\omega_5^2}\sqrt{P_{uud} P_{s\bar{s}}}.
\label{IIGANDwoCM}\eeqna We see that  the direct current
contributions to $G_M^s$ and $G_A^s$, as given in Eqs.
(\ref{IIGMDwoCM}-\ref{IIGADwoCM}), are both negative at $q^2=0$
which contradicts the experiments. However, the transition
current contributions  to $G_M^s$ and $G_A^s$, as given in Eqs.
(\ref{IIGMNDwoCM}-\ref{IIGANDwoCM}) are of opposite sign and in
agreement with the data, independent of the sign of $\delta$. Since
the transition current contributions dominate over the direct
current contributions in the model considered here, it is hence of
interest to see whether we could fit the experimental data of the
proton strangeness form factors with configuration II or some linear
combinations of configurations I and II.

We take the proton and quark masses to be 0.938 and 0.313 GeV,
respectively. The oscillator parameter for the $3q$ core is fixed to
be $\omega_3 = 0.246$ GeV. We then vary the oscillator parameter
$\omega_5$ and the probability $P_{s\bar{s}}$ of the $5q$ component
 $uuds\bar s$,  to fit the experimental data $G_E^{s} + \eta G_M^{s}$
\cite{G0}, which are more directly measured in the experiments and
$G_A^{s}$ as extracted in Ref. \cite{Pate08} where its sign in
low-$q^2$ region is well determined. Both signs of $\delta=\pm 1$
are tried and the best results are then determined.

Our best fits to the experimental data $G_A^{s}$ and
$G_E^{s} + \eta G_M^{s}$, within configuration II, are
shown in Fig. 1 as solid curves, with $\omega_5 = 0.469$ GeV, $P_{s\bar s} =
0.025\%$, and relative phase $\delta_P = +1$, where subscript $P$
refers to the fact that the configuration under consideration has
$\bar s$ in $P$ state. The data denoted with open circles are from
Ref. \cite{G0} and the crosses represent the corresponding values
after the two-boson exchange effects are corrected \cite{Zhou07}. Solid triangles
are the data from HAPPEx \cite{HAPPEX} and open boxes are from Ref. \cite{Pate08}.
The ensuing results for $G_E^s$ and $G_M^s$ are shown in Fig. 2. It
is seen that the agreement with the data are in general quite good
except for $G_E^{s} + \eta G_M^{s}$ and $G_M^s$ at small
values of $q^2$, where there are large experimental uncertainties.

\begin{figure}[htbp]

\begin{center}
\includegraphics[width=0.9\linewidth,angle=0]{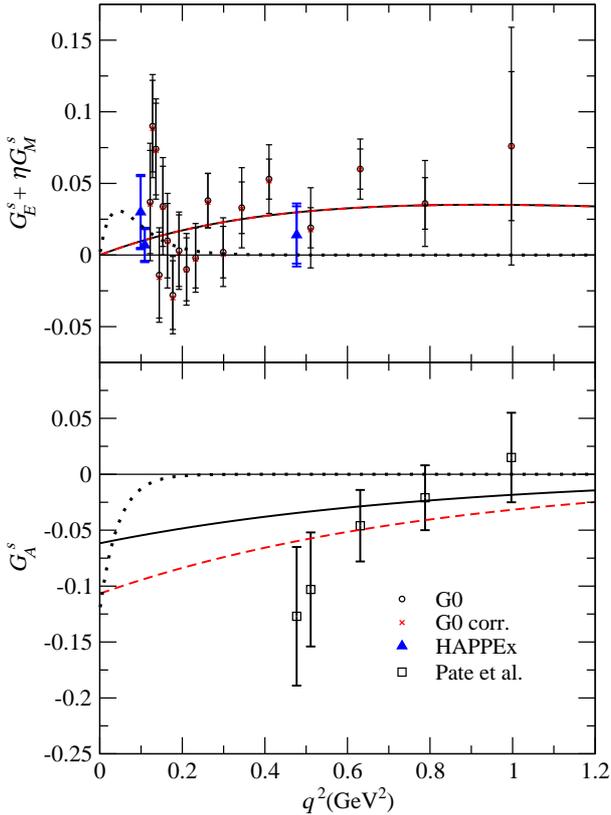}
\end{center}
\caption{Our predictions for form factors $G_E^s +\eta G_M^s$ and
$G_A^s$. The results obtained with configurations with $\bar s$ in
pure $P$ state,  $S$ and $P$ states admixture A, and B are denoted,
respectively, by full, dashed, and dotted lines. Experimental data
from Refs. \cite{G0,HAPPEX,Pate08} are denoted by open circles (G0),
solid triangles (HAPPEx), and open boxes (Pate {\it et al.}),
respectively, while the crosses are the corrected values of
the G0 data by taking into account the two-boson exchange mechanism
\cite{Zhou07}.} \label{FF_1}
\end{figure}

\begin{table*}
\caption{{The strangeness magnetic dipole moment $\mu_s$ (in unit of $\mu_N$),
strangeness contribution to proton spin $\Delta S$, and the radius
of the $uuds\bar s$ component, obtained in our calculation, with
$\bar{s}$ in   pure $P$ state, $S$ and $P$ states admixture A, and B
, as compared with the experiments.}}
\begin{center}
\begin{tabular}{|c|c|c|c|c|c|c|c|c|}
\hline &  & \multicolumn{3}{|c|}{$\bar{s}$ in $S $ and $P$ admixture A $(0.058\%)$} & \multicolumn{3}{|c|}{$\bar{s}$ in $S $ and $P$ admixture B $(2.4\%)$} &
\\ \cline{3-8}
&{$\bar{s}$ in $P$ state} & $S$ state & $P$ state &        & $S$ state & $P$ state &        & Experiments\\
&$(0.025\%)$              & $(8\%)$   & $(92\%)$  & Total  & $(15\%)$  & $(85\%)$  & Total  &  \\

\hline
$\mu_s$                 & $0.066$  & $-0.030$ & $0.096$   & $0.066$     & $-0.80$  & $1.80$   & $1.01$           &  $0.37\pm  0.79$ \cite{Young06}\\
$\Delta S$              & $-0.062$ & $-0.017$ & $-0.090$ & $-0.107$     & $-0.025$ & $-0.097$ & $-0.12$           & $-0.10\pm 0.03$ \cite{EMC}\\\cline{3-8}
$r_{5q}$ & 0.5 fm & \multicolumn{3}{|c|}{0.5 fm} & \multicolumn{3}{|c|}{2.16 fm} & N/A \\

 \hline
\end{tabular}
\end{center} \label{tab:obs_direct}
\end{table*}

\begin{figure}[htbp]

\begin{center}
\includegraphics[width=0.9\linewidth,angle=0]{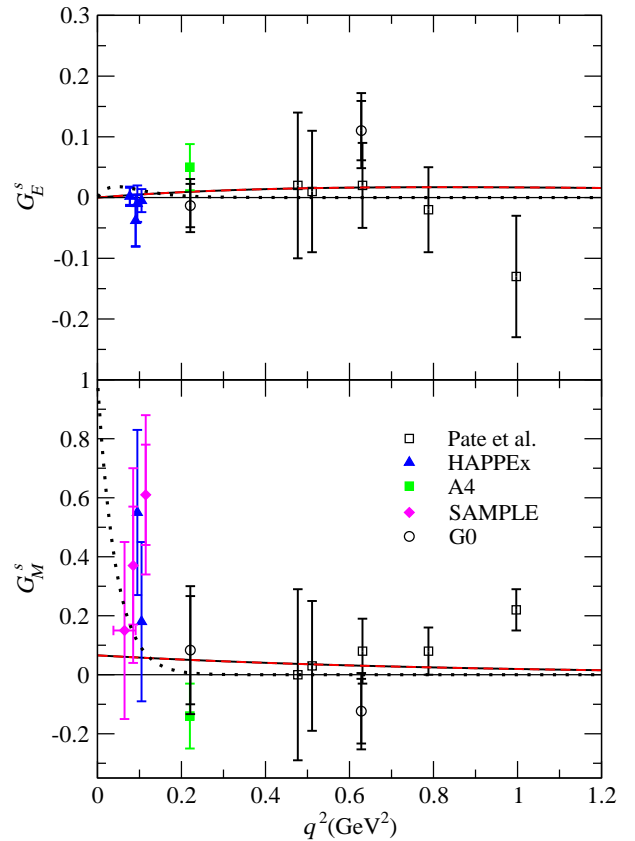}
\end{center}


\caption{Our results for $G_E^s$ and $G_M^s$. Notations are  the
same as in Fig. \ref{FF_1}. Experimental data are from Refs.
\cite{SAMPLE,HAPPEX,A4,G0} and denoted with solid diamonds (SAMPLE),
solid triangles (HAPPEx), solid boxes (A4), and open circles (G0),
respectively.} \label{FF_2}
\end{figure}

We have also explored the possibility of mixing configurations II
and I, namely,
 \beqna |\textrm{proton}\rangle&=&A_3|3q>
 +A_5\sum_{\alpha=S,P}\delta_\alpha b_\alpha|5q;\alpha>,
 \eeqna where
 $|5q;\alpha>$ and $\delta_\alpha$ denote the $uuds\bar s$ states with $\bar s$
 in either $S$ or $P$ states and its relative phase with the
 three-quark state $|3q>$, respectively,
to see whether a better description of the data can be obtained. It
turns out that some improvements can be achieved only for $G_A^s$ at
low-$q^2$ region with a small mixing probability of $b^2_S=8\%$ for
configuration I, relatives phases $\delta_P=1, \delta_S=-1$, and a
combined probability of $P_{s\bar s}=A^2_5=0.058\%$ (called
admixture A), as shown by the dashed curves in Figs. \ref{FF_1} and
\ref{FF_2}.

It is seen that we could fit the data reasonably well with a rather
small probability of $uuds\bar s$ component, e.g., $P_{s\bar s} =
0.025\%\sim 0.058\%$ with either configuration II alone or a mixture
of configurations II and I. It is in sharp contrast to the values of
$P_{s\bar s}=10\sim 15\%$ required in Ref. \cite{RZ06} in order to
fit $G_M^s$. Such a great reduction in $P_{s\bar s}$ needed to
reproduce the experimental data on the strangeness form factors
arises from several sources. These include the correction factor
$\sqrt{3}$ in the evaluation of color matrix elements, change of the
configuration for $\bar s$ from $S$ to $P$ state, removal of the CM
motion of the clusters, and the use of different model parameters.
Each of them enhances the transition current matrix elements by
$\sim 50-120 \%$.  It is interesting to note that our set of
harmonic oscillator model parameters would give rise to a size of
the $uuds\bar s$ to be about 0.5 fm, which is quite close to that
estimated by Ref. \cite{Henley92} using a proton-core-$\phi$ picture
for five-quark system with a scaling factor $s = 1.5$.

The corresponding results for strangeness magnetic moment $\mu_s$,
strangeness contribution to proton spin $\Delta S=G^s_A(0)$, and the
size of $uuds\bar s$ component $r_{5q}$, with $\bar s$ in $P$ state
or admixture of $S$ and $P$ states, are given in Table 1 and
compared with the experiments. The agreement between experiments and
numbers obtained with $\bar s$ in admixture of $S$ and $P$
states seems reasonable.


Recently, Chang and Peng \cite{Chang11} generalize the approach of
Brodsky, Hoyer, Peterson, and Sakai (BHPS) \cite{BHPS80} for the
intrinsic charm quark distribution in the nucleons to the
light-quark sector involving intrinsic $\bar u, \bar d$, and $\bar
s$ sea quarks to analyze the existing $\bar d(x)-\bar u(x),
s(x)+\bar s(x),$ and $\bar u(x)+\bar d(x)-s(x)-\bar s(x)$ data and
conclude that probability $P_{s\bar s}$ for five-quark configuration
$uuds\bar s$ to lie between 0.024-0.029. We explore the consequence
of such a result to our model calculation by fixing $P_{s\bar s} =
2.4\%$ and vary $\omega_5$ to fit the data. The resultant fit we
obtain with $\omega_5 = 0.108$ GeV, which corresponds to a large
size of the five-quark system with $r_{5q}=2.16$ fm, and a small
admixture of $S$ state with a probability of about $15\%$ (called
admixture B), are shown in Figs. \ref{FF_1} and \ref{FF_2} by dotted
lines. The most interesting feature of this fit is the appearance of
a bump in $G^s_E+\eta G^s_M$ in the very low-$q^2$ region with
$q^2\le 0.1  \, \textrm{GeV}^2$, which seems to be hinted by the G0 data but
hampered by large experimental error bars and fluctuations. It would
be worthwhile to carry out experiments in such a low-$q^2$ region if
further theoretical study would support this behavior. All form
factors vanish rapidly beyond $q^2\ge 0.2 \, \textrm{GeV}^2$ because of the
small value of $\omega_5$ and Gaussian nature of the harmonic
oscillator wavefunctions. We  note that the predicted values of
$\mu_s, \Delta S$, and $r_{5q}$, as presented in Table 1, agree with
the data within experimental errors. However, the experimental value of
$0.37\pm 0.79$ was obtained in Ref. \cite{Young06} in a fit
to the form factors including the next-to-leading-order terms of the
$Q^2$-dependence. In another fit with only leading order in $Q^2$ which
has almost same reduced $\chi^2$, a  value
of $\mu_s=0.12\pm 0.55\pm 0.07$ was obtained which would disfavor
configuration B.

In summary, we have reinvestigated, within a nonrelativistic
constituent quark model, the question of whether a five-quark
component with configuration of (4,1) clustering, as previously
considered by Riska and Zou \cite{ZR05,RZ06} can account for the
data of the proton strangeness form factors. Two configurations of
the lowest energies, both consist of four quarks in colored state and
one antiquark are considered. They possess spatial-flavor-spin
symmetry $[31]_X-[4]_{FS}[22]_F[22]_S$ and
$[4]_X-[31]_{FS}[211]_F[22]_S$ ,
  with   antiquark $\bar s$  in the $S$ (configuration I)
  and $P$ states (configuration II), respectively. They are
  degenerate before being splitted by the color hyperfine
quark-quark interaction with configuration II of higher energy.

We have not been able to reproduce the results of Ref. \cite{RZ06}
to substantiate their claim that configuration I is to be preferred
as the dominant configuration of a possible five-quark component.
The claim made in Ref. \cite{RZ06} is based on a particular choice
of the relative phase $\delta$ between $uud$ and $uuds\bar s$
components of the proton wave functions such that $G^s_M$ could be
positive at small $q^2$ region. However, the corrected expression of
$G^s_A$ we obtain for configuration I would lead to a $G^s_A$ to be
of the same sign as $G^s_M$ in the low-$q^2$ region which clearly
contradicts all existing data. We also find that the expression for
the transition current contribution to $G^s_E$ obtained in Ref.
\cite{RZ06} carries an erroneous minus sign.

We then proceed to study configuration II, where $\bar s$ is in $P$
state, and make an effort to remove the CM motions of the clusters
which was not done in Ref. \cite{RZ06}. We demonstrate that it is
possible to give a satisfactory description of the existing data on
the proton strangeness form factors $G^s_E$, $G^s_M$, $G^s_A$, and the
linear combination of $G^s_E + \eta G^s_M$ which is more directly
extracted from the parity-violating asymmetry $A_{PV}$ measured in
elastic electron-proton scattering with a very small probability of
$P_{s\bar s}=0.025\%$ for the $uuds\bar s$ components. The agreement
with $G^s_A$ data can be improved in the low-$q^2$ region by considering an
admixture of configurations I and II with a total $uuds\bar s$
probability $P_{s\bar s}$ increased to $0.058\%$ with configuration
I accounts for $8\%$ of the total.  We further find that
without removing CM motion, $P_{s\bar s}$ would be overestimated by
about a factor of four in the case when transition current dominates.

We have also explored the consequence of a recent claim
\cite{Chang11}, reached from analyzing existing data on $\bar
d(x)-\bar u(x), s(x)+\bar s(x),$ and $\bar u(x)+\bar d(x)-s(x)-\bar
s(x)$, that $P_{s\bar s}$   lies between  $2.4-2.9\%$. A small bump
in both $G^s_E+\eta G^s_M$ and $G^s_E$ in the region of $q^2\le 0.1  \, \textrm{GeV}^2$
for an admixture of configuration I and II.

It is tempting to conclude, from the results presented above, that
the current strangeness form factors experiments seem to indicate that
the dominant configuration in which $uuds\bar s$ arrange themselves
in configuration II, in which $uuds$ is in a completely symmetric
spatial state $[4]_X$ with flavor and spin state symmetry of
$[31]_{FS}[211]_F[22]_S$.   However, there are a few caveats here.
First is that the agreement between our results and the existing
data is not perfect, to say the least. For example, recent data from
A4 at $q^2=0.22  \, \textrm{GeV}^2$ gives a negative value of
$G^s_M=-0.14\pm0.11\pm0.11$.  One might also ask  whether a
nonrelativistic constituent quark model is quantitatively reliable
in evaluating the contributions of transition current which is found
to be dominant in our calculation but is of a relativistic effect in
nature. More study about this issue is clearly needed. Lastly, it is
 known that configurations of $(4,1)$ clustering and $(3,2)$
clustering, like the meson cloud configuration, are not orthogonal.
Then whether $(4,1)$ or $(3,2)$ configuration would be favored might
be just a matter of choosing a different basis if it   turns out
that more precise data will require linear combination of several
$(4,1)$ clustering configurations.

\section*{Acknowledgments}
We acknowledge helpful discussions we have with Drs. Chun-Sheng An,
Fatiha Benmokhtar, Chung-Wen Kao, and BingSong Zou. H.C. Lee
gratefully acknowledges the warm hospitality extended to her during
a brief visit to IHEP, Beijing. This work is supported in part by
the National Science Council of the Republic of China (Taiwan) under
grant No. NSC99-2112-M002-011.

\bibliographystyle{elsarticle-num}

\end{document}